\documentclass[prl,aps,preprint,tightenlines,showpacs,superscriptaddress]{revtex4}
\usepackage{graphics}
\usepackage{color}
\usepackage{epsf}

\def\cp{$CP$\/}
\def\mbc{$M^{}_{\rm bc}$}
\def\mmbc{M^{}_{\rm bc}}
\def\deltaE{$\Delta E$}

\def\phitwo{$\phi^{}_2$}

\def\mevm{~MeV/$c^2$\/}

\def\meve{~MeV}
\def\gevm{~GeV/$c^2$\/}
\def\gevp{~GeV/$c$\/}
\def\geve{~GeV}

\def\ra{\!\rightarrow\!}
\def\bbar{\overline{B}{}^{\,0}}
\def\dbar{\overline{D}{}^{\,0}}

\def\brhorho{$B^0\ra\rho^+\rho^-$}

\def\arhorho{${\cal A}$}
\def\srhorho{${\cal S}$}

\def\cone{\cos\theta^{}_{\!+}}
\def\ctwo{\cos\theta^{}_{\!-}}
\def\conesq{\cos^2\theta^{}_{\!+}}
\def\ctwosq{\cos^2\theta^{}_{\!-}}
\def\sonesq{\sin^2\theta^{}_{\!+}}
\def\stwosq{\sin^2\theta^{}_{\!-}}

\begin{document}

\vspace*{-3\baselineskip}
\resizebox{!}{2.5cm}{\includegraphics{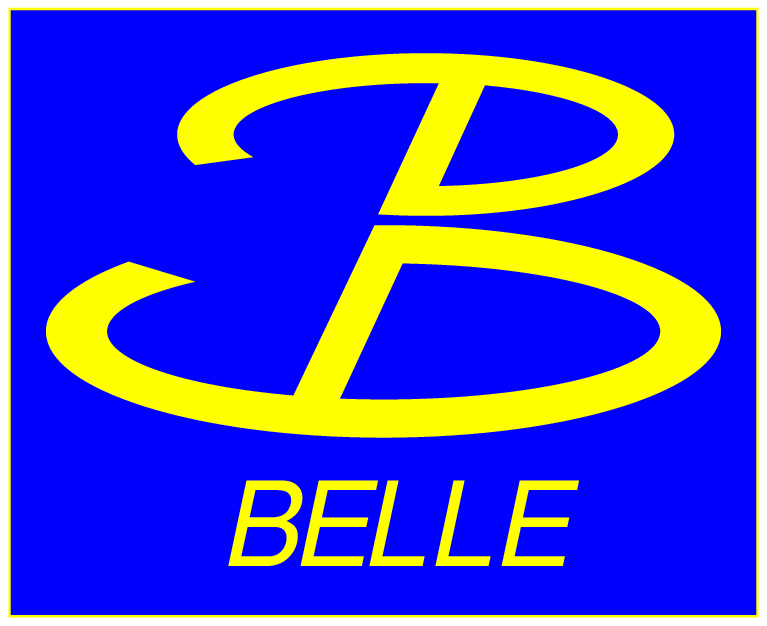}}

\preprint{\vbox{ \hbox{   }
                 \hbox{BELLE Preprint 2006-3}
                 \hbox{KEK Preprint 2005-94}
                 \hbox{UCHEP-06-01}
}}

\title{ 
{\boldmath Measurement of the Branching Fraction, Polarization, 
and \cp\ Asymmetry for \brhorho\ Decays, and Determination of the 
Cabibbo-Kobayashi-Maskawa Phase~\phitwo} }

\affiliation{Budker Institute of Nuclear Physics, Novosibirsk}
\affiliation{Chiba University, Chiba}
\affiliation{Chonnam National University, Kwangju}
\affiliation{University of Cincinnati, Cincinnati, Ohio 45221}
\affiliation{Gyeongsang National University, Chinju}
\affiliation{University of Hawaii, Honolulu, Hawaii 96822}
\affiliation{High Energy Accelerator Research Organization (KEK), Tsukuba}
\affiliation{Hiroshima Institute of Technology, Hiroshima}
\affiliation{Institute of High Energy Physics, Chinese Academy of Sciences, Beijing}
\affiliation{Institute of High Energy Physics, Vienna}
\affiliation{Institute of High Energy Physics, Protvino}
\affiliation{Institute for Theoretical and Experimental Physics, Moscow}
\affiliation{J. Stefan Institute, Ljubljana}
\affiliation{Kanagawa University, Yokohama}
\affiliation{Korea University, Seoul}
\affiliation{Kyungpook National University, Taegu}
\affiliation{Swiss Federal Institute of Technology of Lausanne, EPFL, Lausanne}
\affiliation{University of Ljubljana, Ljubljana}
\affiliation{University of Maribor, Maribor}
\affiliation{University of Melbourne, Victoria}
\affiliation{Nagoya University, Nagoya}
\affiliation{Nara Women's University, Nara}
\affiliation{National Central University, Chung-li}
\affiliation{National United University, Miao Li}
\affiliation{Department of Physics, National Taiwan University, Taipei}
\affiliation{H. Niewodniczanski Institute of Nuclear Physics, Krakow}
\affiliation{Niigata University, Niigata}
\affiliation{Nova Gorica Polytechnic, Nova Gorica}
\affiliation{Osaka City University, Osaka}
\affiliation{Osaka University, Osaka}
\affiliation{Panjab University, Chandigarh}
\affiliation{Peking University, Beijing}
\affiliation{Princeton University, Princeton, New Jersey 08544}
\affiliation{RIKEN BNL Research Center, Upton, New York 11973}
\affiliation{Saga University, Saga}
\affiliation{University of Science and Technology of China, Hefei}
\affiliation{Seoul National University, Seoul}
\affiliation{Sungkyunkwan University, Suwon}
\affiliation{University of Sydney, Sydney, New South Wales}
\affiliation{Tata Institute of Fundamental Research, Bombay}
\affiliation{Toho University, Funabashi}
\affiliation{Tohoku Gakuin University, Tagajo}
\affiliation{Tohoku University, Sendai}
\affiliation{Department of Physics, University of Tokyo, Tokyo}
\affiliation{Tokyo Institute of Technology, Tokyo}
\affiliation{Tokyo Metropolitan University, Tokyo}
\affiliation{Tokyo University of Agriculture and Technology, Tokyo}
\affiliation{University of Tsukuba, Tsukuba}
\affiliation{Virginia Polytechnic Institute and State University, Blacksburg, Virginia 24061}
\affiliation{Yonsei University, Seoul}
   \author{A.~Somov}\affiliation{University of Cincinnati, Cincinnati, Ohio 45221} 
   \author{A.~J.~Schwartz}\affiliation{University of Cincinnati, Cincinnati, Ohio 45221} 
   \author{K.~Abe}\affiliation{High Energy Accelerator Research Organization (KEK), Tsukuba} 
   \author{K.~Abe}\affiliation{Tohoku Gakuin University, Tagajo} 
   \author{I.~Adachi}\affiliation{High Energy Accelerator Research Organization (KEK), Tsukuba} 
   \author{H.~Aihara}\affiliation{Department of Physics, University of Tokyo, Tokyo} 
   \author{D.~Anipko}\affiliation{Budker Institute of Nuclear Physics, Novosibirsk} 
   \author{K.~Arinstein}\affiliation{Budker Institute of Nuclear Physics, Novosibirsk} 
   \author{Y.~Asano}\affiliation{University of Tsukuba, Tsukuba} 
   \author{V.~Aulchenko}\affiliation{Budker Institute of Nuclear Physics, Novosibirsk} 
% ...
\collaboration{The Belle Collaboration}

\noaffiliation

\begin{abstract}
We have measured the branching fraction ${\mathcal B}$, 
longitudinal polarization fraction $f^{}_L$, and \cp\ 
asymmetry coefficients ${\cal A}$ and ${\cal S}$ for \brhorho\ 
decays with the Belle detector at the KEKB $e^+ e^-$ collider
using $253\,{\rm fb}^{-1}$ of data. We obtain 
${\mathcal B}=
\left[\,22.8\,\pm 3.8\,({\rm stat})\,^{+2.3}_{-2.6}\,({\rm syst})\,\right]
\!\times\!10^{-6}$, $f^{}_L =
0.941\,^{+0.034}_{-0.040}\,({\rm stat})\,\pm 0.030\,({\rm syst})$,
${\cal A} = 0.00\,\pm 0.30\,({\rm stat})\,\pm 0.09\,({\rm syst})$, 
and ${\cal S} = 0.08\,\pm 0.41\,({\rm stat})\,\pm 0.09\,({\rm syst})$.
These values are used to constrain the Cabibbo-Kobayashi-Maskawa 
phase~\phitwo; the solution consistent with the Standard Model 
is $\phi^{}_2=(88\,\pm 17)^\circ$ or 
$59^\circ\!<\!\phi^{}_2\!<\!115^\circ$ at 90\%~C.L.
\end{abstract}


\maketitle

{\renewcommand{\thefootnote}{\fnsymbol{footnote}}}
\setcounter{footnote}{0}

One of the main goals of the $e^+e^-$ ``$B$-factories'' is to
determine whether the Cabibbo-Kobayashi-Maskawa~\cite{ckm} 
mixing matrix with three quark generations is unitary; failure 
to satisfy this criterion
would indicate new physics. Unitarity imposes six independent 
constraints upon the matrix elements, one of which is  
$V^*_{ub}V^{}_{ud}+V^*_{cb}V^{}_{cd}+V^*_{tb}V^{}_{td}\!=\!0$.
Plotting this relationship in the complex plane yields a triangle, 
and unitarity is tested by measuring the internal angles 
(denoted $\phi^{}_1,\,\phi^{}_2,\,\phi^{}_3$) to check 
whether they sum to $180^\circ$. The angle $\phi^{}_2$ is 
the phase difference between $V^{*}_{tb}V^{}_{td}$ and
$-V^*_{ub}V^{}_{ud}$ and is measured
via $b\ra u$ decays such as 
$B^0\ra\pi^+\pi^-,\,\rho^\pm\pi^\mp$, and 
$\rho^+\rho^-$~\cite{chargeconjugate}. Of these, 
\brhorho\ gives the most precise value as the
contribution from a possible loop amplitude
(with a different weak phase) is smallest.
The size of the loop amplitude is constrained
by the upper limit on 
${\mathcal B}(B^0\ra\rho^0\rho^0)$~\cite{babar_rho0rho0}.

One determines \phitwo\ by measuring the $\Delta t$ distributions 
of $B^0\bbar$ events, where $\Delta t$ is the difference between 
the decay time of the signal $B^0\ (\bbar)$ and that of the 
opposite-side~$\bbar\ (B^0)$. For $B^0/\bbar\ra\rho^+\rho^-$ 
decays, these distributions have interference terms of opposite 
sign proportional to 
$e^{-|\Delta t|/\tau^{}_B}[\,{\cal A}\cos(\Delta m\,\Delta t)\,+\, 
{\cal S}\sin(\Delta m\,\Delta t)\,]$, where $\Delta m$ is the 
$B^0$-$\bbar$ mass difference and
\arhorho, \srhorho\ are functions of~\phitwo. 
Here we present a measurement of the \brhorho\ branching fraction
${\mathcal B}$, longitudinal polarization fraction $f^{}_L$, and 
coefficients \arhorho\ and~\srhorho, using 
253~fb$^{-1}$ of data recorded by the Belle experiment~\cite{belle_detector} 
at KEKB~\cite{kekb}. 

Candidate \brhorho, $\rho^\pm\ra\pi^\pm\pi^0$ decays 
are selected by requiring two oppositely charged tracks  
satisfying $p^{}_T>0.10$\gevp, $dr<0.2$~cm, and $|dz|<4.0$~cm, 
where $p^{}_T$ is the momentum transverse to the beam axis, 
and $dr$ and $dz$ are the radial and longitudinal distances, 
respectively, between the track and the beam crossing point. 
The tracks are fitted to a common vertex. We require that 
tracks be identified as pions based on information from 
a time-of-flight system, an aerogel \u{C}erenkov counter system,
and the central tracker~\cite{belle_detector}. 
The resulting identification efficiency
is about 89\%, and the kaon 
misidentification rate is about 10\%. Tracks are rejected 
if they satisfy an electron identification criterion based 
on information from an electromagnetic calorimeter~(ECL).

The $\pi^\pm$ candidates are combined with $\pi^0$ candidates 
reconstructed from $\gamma$ pairs having $M^{}_{\gamma\gamma}$ 
in the range 117.8--150.2\mevm\ ($\pm 3\sigma$ in $m^{}_{\pi^0}$ 
resolution). We require $E^{}_\gamma>50~(90)$\meve\ in the ECL 
barrel (endcap), which subtends $32^\circ$--$129^\circ$ 
($17^\circ$--$32^\circ$ and $129^\circ$--$150^\circ$)
with respect to the beam axis. 
To identify $\rho^\pm\ra\pi^\pm\pi^0$ decays, we require 
that $M^{}_{\pi^\pm\pi^0}$ be in the range 0.62--0.92\gevm\ 
($\pm 2\Gamma$ in the $M^{}_{\pi^\pm\pi^0}$ distribution). 
To reduce combinatorial background, the $\pi^0$'s must have 
$p>0.35$\gevp\ in the $e^+e^-$ center-of-mass (CM) frame, and 
$\rho^\pm$ candidates must satisfy 
$-0.80\!<\!\cos\theta^{}_{\!\pm}\!<\!0.98$, 
where $\theta^{}_{\!\pm}$ is the angle between the 
direction of the $\pi^0$ from the $\rho^\pm$ and the 
negative of the $B^0$ momentum in the $\rho^\pm$ rest frame.

To identify \brhorho\ decays, we calculate variables
$\mmbc\!\equiv\!\sqrt{E^2_{\rm beam}-p^2_B}$ and
$\Delta E\!\equiv\!E^{}_B-E^{}_{\rm beam}$, where $E^{}_B$ 
and $p^{}_B$ are the reconstructed energy and momentum 
of the $B$ candidate, and $E^{}_{\rm beam}$ is the beam 
energy, all evaluated in the CM frame. The $\Delta E$ 
distribution has a tail on the lower side due to incomplete 
containment of the electromagnetic shower in the ECL. 
We define a signal region $\mmbc\!\in(5.27,\,5.29)$\gevm\ 
and $\Delta E\!\in(-0.12,\,0.08)$\geve.

We determine whether a $B^0$ or $\bbar$ evolved and 
decayed to $\rho^+\rho^-$ by tagging the $b$ flavor of 
the non-signal (opposite-side) $B$ decay in the event. 
This is done using a tagging algorithm~\cite{tagging} 
that categorizes charged leptons, kaons, and $\Lambda$'s 
found in the event. The algorithm returns two parameters: 
$q$, which equals $+1\,(-1)$ when the opposite-side $B$
is most-likely a $B^0\,(\bbar)$; and $r$, which indicates the 
tag quality as determined from Monte Carlo (MC) simulation
and varies from $r\!=\!0$ for no flavor discrimination to 
$r\!=\!1$ for unambiguous flavor assignment. 

The dominant background is $e^+e^-\!\ra q\bar{q}\ (q=u,d,s,c)$
production. We discriminate against this using event topology: 
$e^+e^-\!\ra q\bar{q}$ events tend to be jet-like
in the CM frame, while $e^+e^-\!\ra B\overline{B}$ 
tends to be spherical. To quantify sphericity, we 
calculate 16 modified Fox-Wolfram moments and combine them into 
a Fisher discriminant~\cite{KSFW}. We calculate a probability 
density function (PDF) for this discriminant and multiply it 
by a PDF for $\cos\theta^{}_B$, where $\theta^{}_B$ is the 
polar angle in the CM frame between the $B$ direction and 
the beam axis. $B\overline{B}$ events have a 
$1\!-\!\cos^2\theta^{}_B$ distribution while $q\bar{q}$ 
events tend to be uniform in $\cos\theta^{}_B$. The PDFs 
for signal and $q\bar{q}$ are obtained from MC simulation 
and a sideband [$\mmbc\!\in(5.21,\,5.26)$\gevm], respectively. 
These PDFs are used to calculate a 
signal likelihood ${\cal L}^{}_s$ and $q\bar{q}$ likelihood 
${\cal L}^{}_{q\bar{q}}$, and we require that 
${\cal R} ={\cal L}^{}_s/({\cal L}^{}_s + {\cal L}^{}_{q\bar{q}})$
be above a threshold. 
As the tagging parameter $r$ also discriminates against 
$q\bar{q}$ events, we divide the data into six $r$ 
intervals (denoted $\ell\!=\!1\!-\!6$) and determine the 
${\cal R}$ threshold separately for each. 

The overall efficiency (from MC simulation) is $(3.19\pm 0.02)$\%.
This value corresponds to $f^{}_L\!=\!1$; 
the change in efficiency (+5.0\%) for $f^{}_L$ equal to its 
central value measured below is taken as a systematic error.
The fraction of events having multiple candidates is 9.5\%;
most of these arise from fake $\pi^0$'s combining with good 
tracks, and thus we choose the best candidate based on 
$|M^{}_{\gamma\gamma}\!-\!m^{}_{\pi^0}|$. In MC simulation 
this correctly identifies the \brhorho\ decay about 90\% of 
the time. A small fraction of signal decays (5.7\% for longitudinal
polarization) have $\geq$\,1 $\pi^\pm$ daughters incorrectly 
identified but pass all selection criteria; these are referred to
as ``self-cross-feed''~(SCF) events. Their vertex
positions (and hence $\Delta t$ values) are smeared. 

We determine the signal yield using two unbinned 
maximum likelihood (ML) fits. We first fit the 
\mbc-\deltaE\ distribution in the wide range
$\mmbc\!\in(5.21,\,5.29)$\gevm\ and
$\Delta E\!\in(-0.20,\,0.30)$\geve\ 
to obtain the 
$B^0\ra(\rho^+\rho^-\!+\!\mbox{\rm nonresonant})$ 
yield $N^{}_{(\rho\rho\,+ {\rm nr})}$; we then fit 
the $M^{}_{\pi^\pm\pi^0}$ distribution of events 
in the \mbc-\deltaE\ signal region
to obtain the nonresonant 
$\rho^\pm\pi^\mp\pi^0 + \pi^\pm\pi^\mp\pi^0\pi^0$
fraction.

\begin{figure}[t]
\mbox{\epsfxsize=5.75in \epsfbox{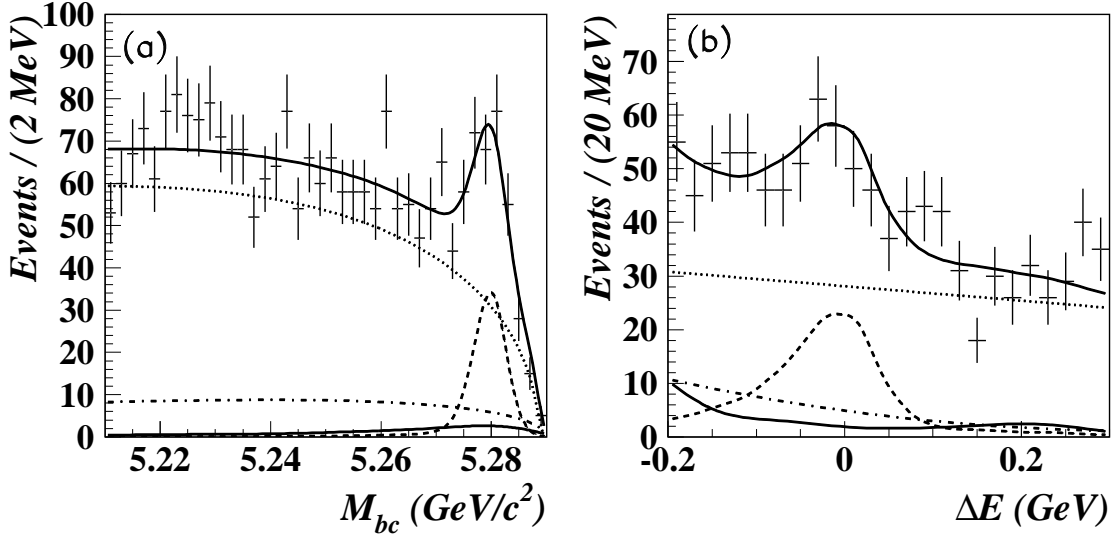}}
\caption{(a)~\mbc\ for events with
$\Delta E\!\in(-0.10,\,0.06)$\geve.
(b)~\deltaE\ for 
$\mmbc\!\in(5.27,\,5.29)$\gevm. The curves 
show fit projections: the dashed curve is $\rho^+\rho^-\!+\rho\pi\pi$,
the dotted curve is $q\bar{q}$, the dot-dashed curve is $b\ra c$, 
the small solid curve is $b\ra u$, and the large solid curve
is the total.} 
\label{fig:one}
\end{figure}

For the first fit we include PDFs for 
signal $\rho^+\rho^-$ and $b\ra c$, $b\ra u$,
and $q\bar{q}$ backgrounds. The PDFs for signal 
and $b\ra u$ are two-dimensional distributions 
obtained from MC simulation; the PDF for $b\ra c$ is the 
product of a threshold (``ARGUS''~\cite{argus}) 
function for \mbc\ and a quadratic polynomial for 
\deltaE, also obtained from MC simulation. The PDF for 
$q\bar{q}$ is an ARGUS function for \mbc\ and a linear function 
for \deltaE; the latter's slope depends on the tag 
quality bin~$\ell$. All $q\bar{q}$ shape parameters are 
floated in the fit. The $b\ra u$ background is dominated by 
$B\ra (\rho\,\pi,\,a^{}_1\pi,\,a^{}_1\,\rho)$ decays; 
as their contributions are small, their normalization is 
fixed to that from MC simulation. For $B^+\ra a^+_1\pi^0$ and 
$B\ra a^{}_1\,\rho$ modes, the branching fractions (unmeasured) 
used in the simulations are $3\times 10^{-5}$ and $2\times 10^{-5}$, 
respectively; we vary these by $\pm 50$\% and $\pm 100$\%, respectively, 
to obtain the systematic error due to these estimates. 
The result of the fit is 
$N^{}_{(\rho\rho\,+ {\rm nr})}\!\!=207\,^{+28}_{-29}$ 
events. Figure~\ref{fig:one} shows the final event sample 
and projections of the fit.

\begin{figure}[t]
\mbox{\epsfxsize=5.75in \epsfbox{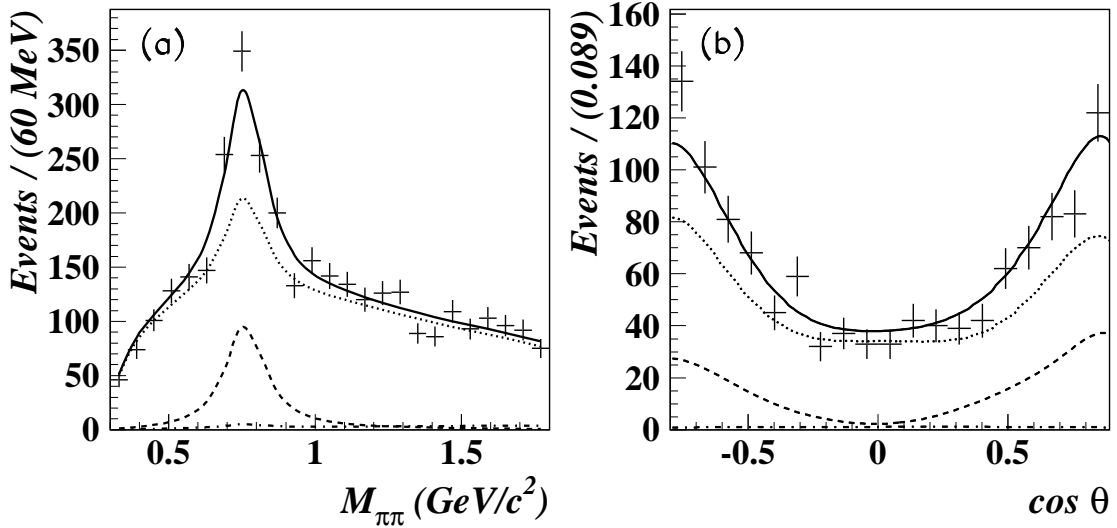}}
\caption{(a)~$M^{}_{\pi^\pm\pi^0}$
for events in the \mbc-\deltaE\ signal region that satisfy
$M^{}_{\pi\pi^0}{\rm\,(not\ fit)}\!\in(0.62,\,0.92)$\gevm.
(b)~Sum of $\cos\theta^{}_{\pm}$ distributions for 
events in the signal region that satisfy 
$M^{}_{\pi^\pm\pi^0}{\rm\,(both)}\!\in(0.62,\,0.92)$\gevm.
The curves show fit projections: 
the dashed curve is $\rho^+\rho^-$,
the dot-dashed curve is $\rho\pi\pi$, 
the dotted curve is $q\bar{q}\,+(b\ra c)+(b\ra u)$, 
and the solid curve is the total.} 
\label{fig:two}
\end{figure}

For the subsequent fit, we require that events 
be in the \mbc-\deltaE\ signal region 
and fit $M^{}_{\pi^\pm\pi^0}$ in the wide range 
0.30--1.80\gevm. One $\rho$ candidate is required to satisfy 
$M^{}_{\pi\pi^0}\!\in(0.62,\,0.92)$\gevm;
the mass of the other $\rho$ candidate is then fit.
We include additional PDFs for 
nonresonant $B\ra\rho\pi\pi$ and $B\ra\pi\pi\pi\pi$ decays; 
these are taken from MC simulation assuming three- 
and four-body phase space distributions.
However, the fit result for $\pi\pi\pi\pi$
is $\ll$\,1\% and thus we set this fraction to zero. 
The PDFs for $\rho^+\rho^-$ and $b\ra u$ are also
taken from MC simulation. The PDFs for $b\ra c$ and $q\bar{q}$
are combined and taken from the sideband
$\mmbc\!\in(5.22,\,5.26)$\gevm; 
we check with MC simulation that the shapes of these backgrounds 
and their ratio in the sideband region are close to those 
in the signal region. We impose the constraint that the 
fraction of $(\rho^+\rho^-\!+\rho\pi\pi)$ events
in the $M^{}_{\pi^\pm\pi^0}$ range 0.62--0.92\gevm\ 
equals that obtained from the \mbc-\deltaE\ fit;
there is then only one free parameter. The fit 
obtains $\tilde{f}^{}_{\rho\pi\pi} \equiv
f^{}_{\rho\pi\pi}/(f^{}_{\rho\rho}+f^{}_{\rho\pi\pi})
=(6.3\,\pm 6.7)\%$, and thus $N^{}_{\rho\rho}=
(1-\tilde{f}^{}_{\rho\pi\pi}) N^{}_{(\rho\rho\,+ {\rm nr})}
=194\,\pm 32$, where the error is statistical and obtained
from a ``toy'' MC study (since the errors on 
$\tilde{f}^{}_{\rho\pi\pi}$ and $N^{}_{(\rho\rho\,+ {\rm nr})}$
are correlated). 
This value agrees well with the $\rho^+\rho^-$ yield obtained 
from the $M^{}_{\pi\pi^0}$ fit~(141 events) multiplied by the 
ratio of acceptances~(1.33).
Figure~\ref{fig:two}(a) shows the data and 
projections of the fit.

The branching fraction is $N^{}_{\rho\rho}/
(\varepsilon\cdot\varepsilon^{}_\pi\cdot N^{}_{B\overline{B}})$,
where $N^{}_{\rho\rho}$ is the number of \brhorho\ candidates,
$N^{}_{B\overline{B}}$ is the number of $B\overline{B}$ pairs
produced [$(274.8\,\pm 3.1)\times 10^6$], 
$\varepsilon$ is the acceptance and event 
selection efficiency obtained from MC simulation, and 
$\varepsilon^{}_\pi$ is a correction factor for 
the $\pi^\pm$ identification requirement to account for small 
differences between data and the simulation ($0.969\pm 0.012$). 
The result is ${\mathcal B}=(22.8\pm 3.8)\times 10^{-6}$, where 
the error is statistical.

There are eleven main sources of systematic error. These are
typically evaluated by varying the relevant parameter(s) by 
$1\sigma$ and noting the change in~${\mathcal B}$. The sources are: 
track reconstruction efficiency (1.2\% per track);
$\pi^0$ efficiency (4\% per $\pi^0$);
calibration factors (obtained from a large 
$B^+\ra\dbar\rho^+\ra K^+\pi^-\pi^0\rho^+$ sample) 
used to correct the signal \mbc-\deltaE\,\ PDF to better match the data;
the \mbc-\deltaE\ shapes for $b\ra c$; 
the fraction and \mbc-\deltaE\ shapes for $b\ra u$;
the \deltaE\ range fit; 
statistics of the MC simulation used to calculate $\varepsilon$; 
the dependence of $\varepsilon$ upon the polarization;
uncertainties in $\varepsilon^{}_\pi$ and $N^{}_{B\overline{B}}$;
and the $q\bar{q}$ suppression requirement.
Combining these in quadrature gives a total 
systematic error of $+10.1\%$ and $-11.6\%$. Thus,
\begin{eqnarray}
\hspace*{-0.14in}
{\mathcal B}^{}_{B\rightarrow\rho^+\rho^-}\!\! & = &\!\!
\left[\,22.8\,\pm 3.8\,({\rm stat})\,^{+2.3}_{-2.6}\,({\rm syst})\,\right]
\times 10^{-6}\,.
\label{eqn:result}
\end{eqnarray}

To determine the longitudinal polarization fraction $f^{}_L$, 
we perform an unbinned ML fit to the $\theta^{}_+,\theta^{}_-$
helicity angle distribution. This distribution is proportional to 
$\left[\,4f^{}_L\conesq\ctwosq +(1-f^{}_L)\sonesq\stwosq\,\right]$. 
In the fit, this PDF is multiplied by an acceptance function 
determined from MC simulation. The acceptance is modeled as 
the product $A(\cone)\cdot A(\ctwo)$, where $A$ is a polynomial.

We fit events in the \mbc-\deltaE\ signal region that satisfy
$M^{}_{\pi^\pm\pi^0}\!\in(0.62,\,0.92)$\gevm. 
We include PDFs for signal, $\rho\pi\pi$, and $b\ra c$, 
$b\ra u$, and $q\bar{q}$ backgrounds. The PDFs for $b\ra c$ 
and $q\bar{q}$ are combined and determined from the sideband
$\mmbc\!\in(5.21,\,5.26)$\gevm, $\Delta E\!\in(-0.12,\,0.12)$\geve;
we check with MC simulation that the shapes of these backgrounds 
and their ratio in the sideband region are close to 
those in the signal region.
The PDF for $b\ra u$ is taken from MC simulation. 
The fraction of $\rho^+\rho^-\!+\rho\pi\pi$ 
is taken from the \mbc-\deltaE\ fit; the component 
$f^{}_{\rho\pi\pi}$ alone is taken from the $M^{}_{\pi^\pm\pi^0}$ 
fit. The fraction of $b\ra u$ background is small and taken
from MC simulation. Since $f^{}_{(q\bar{q}\,+\,b\rightarrow c)}\!=\!
1\!-\!f^{}_{\rho\rho}\!-\!f^{}_{\rho\pi\pi}\!-\!f^{}_{b\rightarrow u}$,
there is only one free parameter in the fit.
The result is $f^{}_L =0.941\,^{+0.034}_{-0.040}$,
where the error is statistical. Figure~\ref{fig:two}(b) 
shows the data and projections of the fit. 

There are eight main sources of systematic error in~$f^{}_L$:
the $\rho^+\rho^-\!\!+\rho\pi\pi$ fraction $(+0.013, -0.012)$;
the $\rho\pi\pi$ component alone $(+0.021, -0.020)$;
the pion identification efficiency, which affects the 
acceptance $(+0.000, -0.004)$;
misreconstructed \brhorho\ decays $(+0.005, -0.000)$;
the $q\bar{q}$ suppression requirement $(\pm 0.013)$; 
interference of longitudinally polarized $\rho$'s with 
an $S$-wave $\pi^\pm\pi^0$ system in 
$B^0\ra\rho\pi\pi$ $(+0.003, -0.005)$;  
a very small bias in the fitting procedure measured from
a large toy MC sample $(+0.000, -0.005)$; and uncertainty 
in the $q\bar{q}\,+(b\ra c)$ background shape $(+0.004, -0.014)$. 
This last uncertainty is evaluated by taking the background
shape from alternative \mbc, \deltaE\ sidebands. Combining 
all errors in quadrature gives a total systematic error of 
$\pm0.030$.~Thus,
\begin{eqnarray}
f^{}_L & = &
0.941\,^{+0.034}_{-0.040}\,({\rm stat})\,\pm 0.030\,({\rm syst})\,.
\label{eqn:pol_result}
\end{eqnarray}

To determine \cp\ coefficients \arhorho\ and \srhorho, we 
divide the data into $q\!=\!\pm 1$ tagged subsamples and 
do an unbinned ML fit to their $\Delta t$ distributions. 
Since $B^0$'s and $\bbar$'s are approximately at rest in 
the $\Upsilon(4S)$ frame, and the $\Upsilon(4S)$ travels 
with $\beta\gamma=0.425$ nearly along the beam axis~($z$), 
$\Delta t$ is determined from the $z$ displacement between 
the $\rho^+\rho^-$ and tag-side decay vertices:
$\Delta t \approx (z_{CP} - z_{\rm tag})/\beta\gamma c$.

The likelihood function for event $i$ is a sum of terms: 
\begin{eqnarray*} 
{\cal L}^{}_i & = & 
f^{(i)}_{\rho\rho}\,{\cal P}(\Delta t)^{}_{\rho\rho} +
f^{(i)}_{\rm SCF}\,{\cal P}(\Delta t)^{}_{\rm SCF} +
f^{(i)}_{\rho\pi\pi}\,{\cal P}(\Delta t)^{}_{\rho\pi\pi} +  \\ 
 & & \hskip0.50in
f^{(i)}_{b\rightarrow c}\,{\cal P}(\Delta t)^{}_{b\rightarrow c} +  
f^{(i)}_{b\rightarrow u}\,{\cal P}(\Delta t)^{}_{b\rightarrow u} +  
f^{(i)}_{q\bar{q}}\,{\cal P}(\Delta t)^{}_{q\bar{q}}\,, 
\end{eqnarray*}
where the weights $f^{(i)}$ are functions of \mbc\ and \deltaE\
and are normalized to the event fractions 
obtained from the \mbc-\deltaE\ and $M^{}_{\pi^\pm\pi^0}$ fits. 
The PDFs ${\cal P}(\Delta t)$ are obtained from MC simulation
for $b\ra c$ and $b\ra u$ and from an \mbc\ sideband for $q\bar{q}$.
We include a term for SCF events in which a $\pi^\pm$ daughter 
is swapped with a tag-side track; the PDF and function 
$f^{}_{\rm SCF}$ are also obtained from MC simulation.

The signal PDF is 
$e^{-|\Delta t|/\tau^{}_{B^0}}/(4\tau^{}_{B^0})\times
\left\{1\mp \Delta\omega^{}_{\ell}\pm (1-2\omega^{}_{\ell}) 
\left[\,{\cal A}\cos(\Delta m\,\Delta t) +
{\cal S}\sin(\Delta m\,\Delta t)\,\right]\right\}$,
where the upper (lower) sign corresponds to $B^0 (\bbar)$ tags,
$\omega^{}_{\ell}$ is the mistag fraction for the $\ell$th bin of 
tagging parameter~$r$, and $\Delta\omega^{}_\ell$ is a possible 
difference in $\omega^{}_\ell$ between $B^0$ and $\bbar$ tags. 
Values of $\omega^{}_\ell$ and $\Delta\omega^{}_\ell$ are 
determined from a large $B^0\ra D^{*-}\ell^+\nu$ sample.
Coefficients ${\cal A}$ and ${\cal S}$ receive contributions 
from longitudinally~($L$) and transversely~($T$) polarized 
amplitudes, e.g., 
${\cal A}\!=\!f^{}_L{\cal A}^{}_L+(1\!-\!f^{}_L){\cal A}^{}_T$. 
The transversely polarized amplitude has a $CP$-odd component. 
For a negligible penguin contribution,~${\cal A}^{}_T\!=\!{\cal A}^{}_L$ but
${\cal S}^{}_T\!=\![(1\!-\!f^{}_L\!-\!2f^{}_{CP \mbox{-}{\rm odd}})/(1-f^{}_L)]
\,{\cal S}^{}_L$\/;~since
$f^{}_{CP \mbox{-}{\rm odd}}\leq f^{}_T$ and $f^{}_T$ 
is small, we assume ${\cal A}\!=\!{\cal A}^{}_L$,
${\cal S}\!=\!{\cal S}^{}_L$, and take the possible 
difference as a systematic error.

\begin{figure*}[t]
\hbox{\hskip0.20in
\mbox{\epsfxsize=2.75in \epsfbox{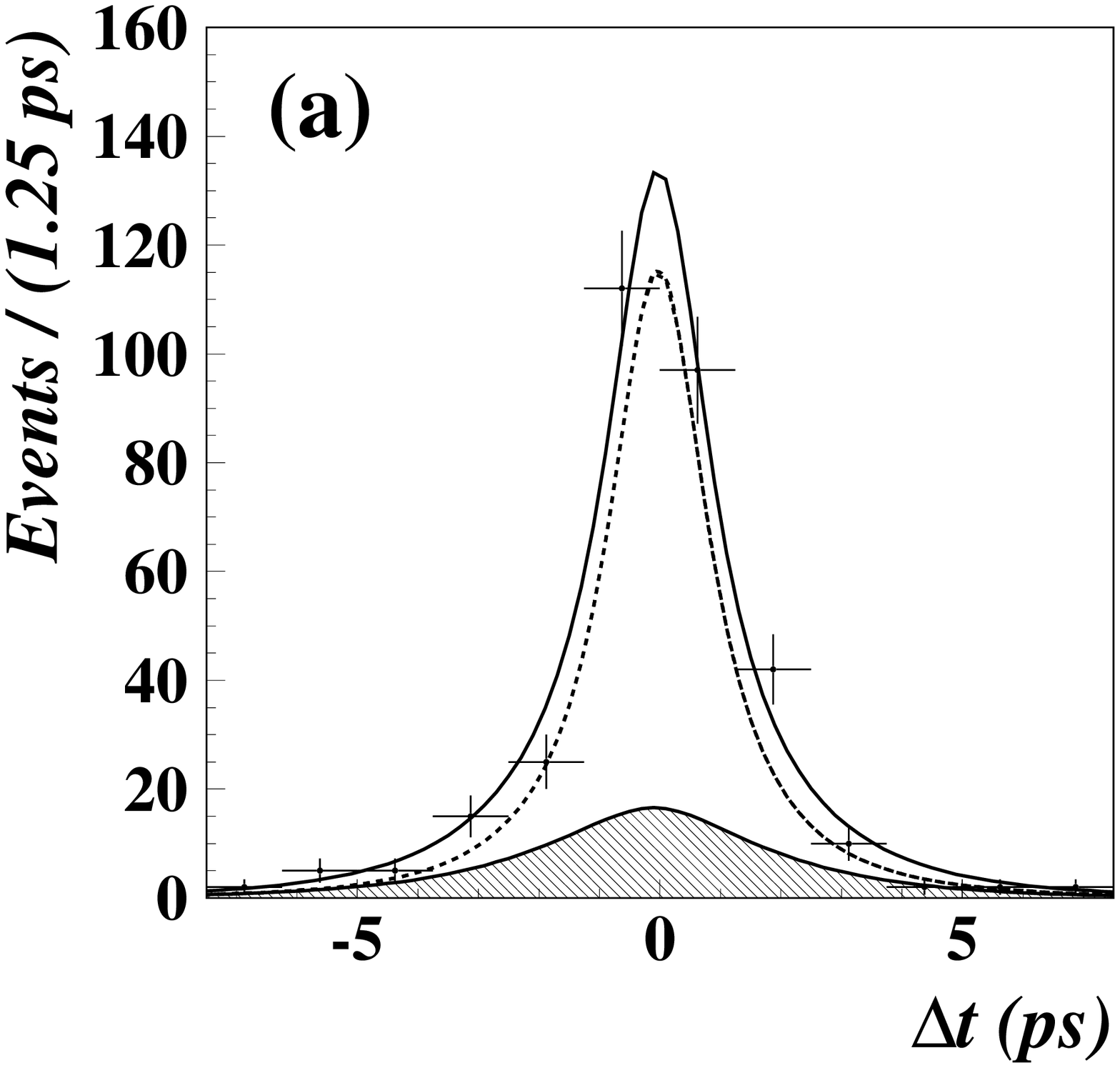}}
\hskip0.30in
\mbox{\epsfxsize=2.75in \epsfbox{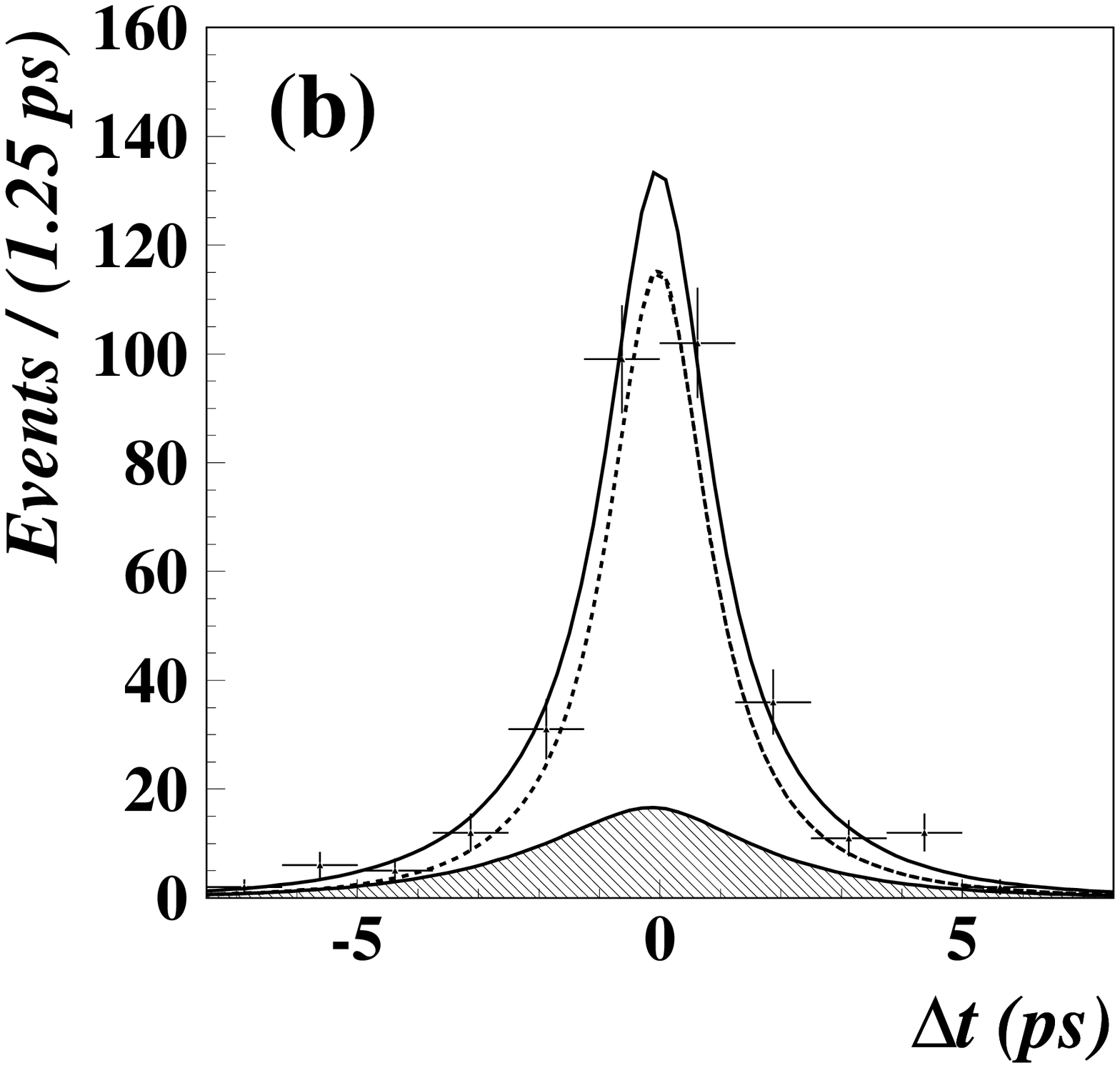}}}
\vskip0.20in
\hbox{\hskip0.20in
\mbox{\epsfxsize=2.75in \epsfbox{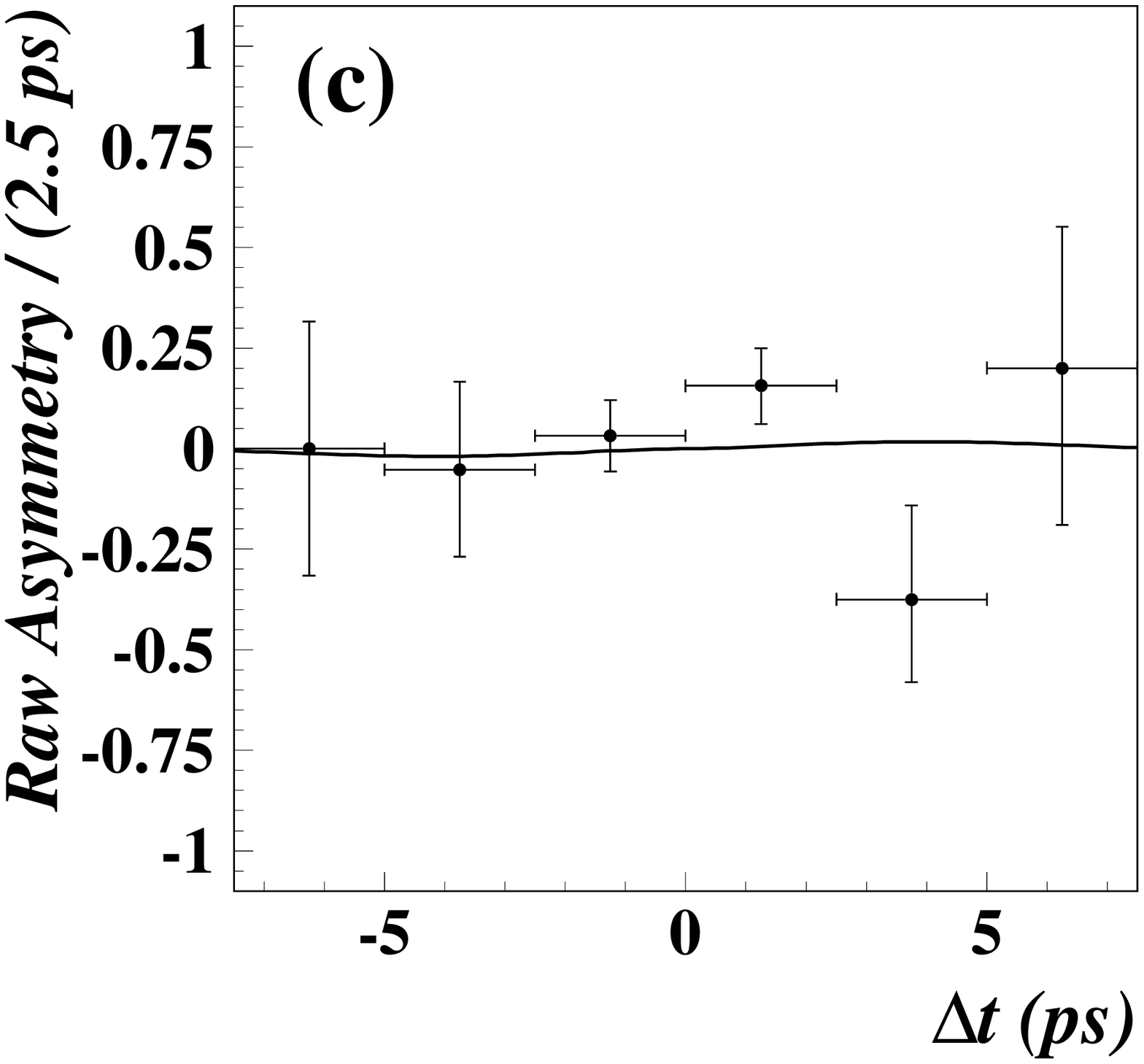}}
\hskip0.30in
\mbox{\epsfxsize=2.75in \epsfbox{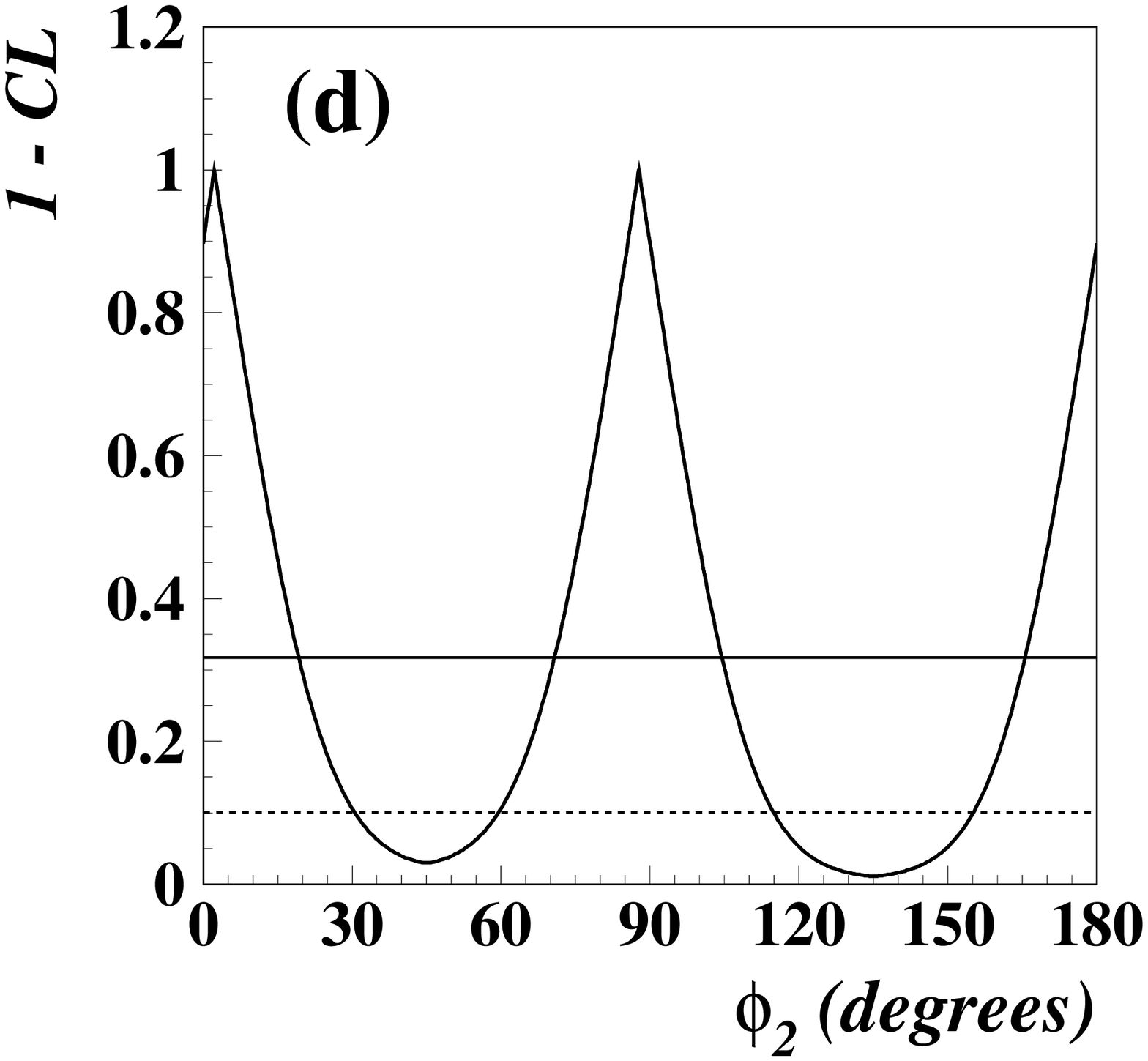}}
}
\caption{The $\Delta t$ distribution of events 
in the \mbc-\deltaE\ signal region that satisfy
$M^{}_{\pi^\pm\pi^0}\!\in(-0.62,\,0.92)$\gevm, 
and projections of the fit.
(a)~$q\!=\!+1$ tags;
(b)~$q\!=\!-1$ tags;
(c)~raw \cp\ asymmetry for good tags ($0.5\!<\!r\!<\!1.0$);
(d)~$1-{\rm C.L.}$ vs.\ \phitwo. 
In (a) and (b), the hatched region (dashed line) shows 
signal (background) events. In (d), the solid (dashed) 
horizontal line denotes ${\rm C.L.}\!=\!68.3$\%~(90\%). }
\label{fig:three}
\end{figure*}

The signal PDF is convolved with the same $\Delta t$
resolution function as that used for Belle's 
$\sin2\phi^{}_1$ measurement~\cite{resolution}.
The PDFs ${\cal P}^{}_{\rho\pi\pi}$ and ${\cal P}^{}_{\rm SCF}$ 
are exponential with $\tau\!=\!\tau^{}_B$ and 
$\tau\!\approx\!0.93$~ps (from MC simulation), respectively; 
these are smeared by a common resolution function. 
We determine \arhorho\ and \srhorho\ by maximizing
$\sum_i \log{\cal L}^{}_i$, where $i$ runs over the 
656 events in the \mbc-\deltaE\ signal region that 
satisfy $M^{}_{\pi^\pm\pi^0}\!\in(0.62,\,0.92)$\gevm. 
The results are ${\cal A}\!=\!0.00\,\pm 0.30$
and ${\cal S}\!=\!0.08\,\pm 0.41$, where the errors 
are statistical. The correlation coefficient is~$-0.057$.
These values are consistent with no \cp\ violation 
(${\cal A}={\cal S}=\!0$); the errors are consistent 
with expectations based on MC simulation. Figure~\ref{fig:three} 
shows the data and projections of the fit.

The sources of systematic error are 
listed in Table~\ref{tab:cpv_systematics}. 
The error due to wrong-tag fractions is evaluated by
varying $\omega^{}_\ell$ and $\Delta\omega^{}_\ell$ values.
The effect of a possible asymmetry in $b\ra c$ and $q\bar{q}$ 
is evaluated by adding such an asymmetry to the $b\ra c$ and 
$q\bar{q}$ $\Delta t$ distributions. 
The error due to transverse polarization
is obtained by first setting $f^{}_L$ equal to its central value 
and varying ${\cal A}^{}_T$, ${\cal S}^{}_T$ from $-1$ to $+1$; 
then assuming ${\cal A}^{}_T\!=\!{\cal A}^{}_L$,
${\cal S}^{}_T\!=\!-{\cal S}^{}_L$ 
($f^{}_T$ is \cp-odd), and varying $f^{}_L$ by its error. 
The sum in quadrature of all systematic errors is
$\pm\,0.09$.~Thus,
\begin{eqnarray}
{\cal A}^{}_L & = &
0.00\,\pm 0.30\,({\rm stat})\,\pm 0.09\,({\rm syst}) \\
{\cal S}^{}_L & = &
0.08\,\pm 0.41\,({\rm stat})\,\pm 0.09\,({\rm syst})\,.
\label{eqn:cpv_result}
\end{eqnarray}
These values are similar to those obtained by BaBar~\cite{babar_alpha}.

\begin{table}[htb]
\caption{Systematic errors for \cp\ coefficients ${\cal A}$ and ${\cal S}$.}
\label{tab:cpv_systematics}
\vskip0.20in
\begin{tabular}{|l|c|c|c|c|}
\hline
{\bf Type} & \multicolumn{2}{|c|}{\boldmath $\Delta {\cal A}$ ($\times 10^{-2}$)}   
     & \multicolumn{2}{|c|}{\boldmath $\Delta {\cal S}$ ($\times 10^{-2}$)} \\\cline{2-5}
            &  {\boldmath $+\sigma$} & {\boldmath $-\sigma$} & 
{\boldmath $+\sigma$} & {\boldmath $-\sigma$} \\
\hline

Wrong tag fractions   &  0.5  &  0.6   &   0.8  &  0.8  \\
Parameters $\Delta m,\,\tau^{}_{B^0}$  
                      &  0.1  &  0.1   &   0.9  &  0.9  \\
Resolution function   &  1.3  &  1.3   &   1.3  &  1.3  \\
Background $\Delta t$ distributions    &   1.6  &  1.5  &  2.3  & 2.5 \\
Component fractions       &  2.1  &  2.6   &   5.1  &  4.5  \\
Background asymmetry  &  0.0      &  2.0   &   0.0       &  4.3  \\
Possible fitting bias    &  0.0      &  1.0   &    0.7  &  0.0   \\        
Vertexing             &  4.1  &  2.8   &    1.3  &  1.4   \\
Tag-side interference~\cite{Long} &  3.7 &  3.7 &   0.1  & 0.1 \\
Transverse polarization   & 6.3  & 6.3  & 7.1  & 5.8    \\
\hline
{\bf Total}  &  $+8.9$\ \  & $-8.8$\ \  &  $+9.3$\ \  &  $-9.2$\ \  \\  
\hline
\end{tabular}
\vskip0.30in
\end{table}

We use these values and the branching fractions for 
$B^0\ra\rho^+\rho^-$\,\cite{br_babar}, $\rho^+\rho^0$\,\cite{pdg}, 
and $\rho^0\rho^0$\,\cite{babar_rho0rho0} to constrain~\phitwo. 
We assume isospin symmetry~\cite{gronau_london} and follow 
Ref.~\cite{charles}, neglecting a possible $I\!=\!1$ contribution 
to \brhorho~\cite{falk}. 
We first fit the measured values to obtain a minimum $\chi^2$ 
(denoted $\chi^2_{\rm min}$); we then scan \phitwo\ from 
0$^\circ$ to 180$^\circ$, calculating the difference 
$\Delta\chi^2\equiv\chi^2(\phi^{}_2)-\chi^2_{\rm min}$. We insert 
$\Delta\chi^2$ into the cumulative distribution function for 
the $\chi^2$ distribution for one degree of freedom to obtain 
a confidence level (C.L.) for each \phitwo\ value. 
The resulting function $1\!-\!{\rm C.L.}$ [Fig.~\ref{fig:three}(d)] 
has more than one peak due to ambiguities that arise
when solving for $\phi^{}_2$. However, only one solution
is consistent with the Standard Model~\cite{pdg}:
$(88\,\pm 17)^\circ$ or
$59^\circ\!<\!\phi^{}_2\!<\!115^\circ$ at 90\%~C.L.

In summary, using 253~fb$^{-1}$ of data we have measured the 
branching fraction, polarization 
fraction, and \cp\ coefficients \arhorho\ and \srhorho\ for
\brhorho\ decays, and constrained the angle~$\phi^{}_2$.

We thank the KEKB group for the excellent
operation of the accelerator, the KEK cryogenics
group for the efficient operation of the solenoid,
and the KEK computer group and the NII for valuable computing and
Super-SINET network support.  We acknowledge support 
from MEXT and JSPS (Japan); ARC and DEST (Australia); NSFC (contract
No.~10175071, China); DST (India); the BK21 program of MOEHRD and the
CHEP SRC program of KOSEF (Korea); KBN (contract No.~2P03B 01324,
Poland); MIST (Russia); MESS (Slovenia); NSC and MOE (Taiwan); 
and DOE (USA).


\begin{thebibliography}{99}
\bibitem{chargeconjugate} 
Charge-conjugate modes are included throughout this paper
unless noted otherwise.
\end{thebibliography}
\end{document}